\documentclass{aastex6}
\pdfoutput=1 

\usepackage{graphicx}
\usepackage{comment}

\usepackage{natbib}

\begin{document}

\title{
 Shining Light on Quantum Gravity with Pulsar--Black Hole Binaries
}

\author{John Estes,\altaffilmark{1} Michael Kavic,\altaffilmark{2} Matthew Lippert\altaffilmark{3}}
\affil{Department of Physics, \\Long Island University,\\Brooklyn, NY 11201, United States}
\and
\author{John H.~Simonetti\altaffilmark{4}}
\affil{Department of Physics, \\Virginia Tech, \\Blacksburg, VA, United States}

\altaffiltext{1}{John.Estes@liu.edu}
\altaffiltext{2}{Michael.Kavic@liu.edu}
\altaffiltext{3}{Matthew.Lippert@liu.edu (corresponding author)}
\altaffiltext{4}{jhs@vt.edu}
\begin{abstract}

Pulsars are some of the most accurate clocks found in nature, while black holes offer a unique arena for the study of quantum gravity.  As such, pulsar--black hole (PSR--BH) binaries provide ideal astrophysical systems for detecting the effects of quantum gravity.  With the success of aLIGO and the advent of instruments like the SKA and eLISA, the prospects for the discovery of such PSR--BH binaries are very promising.  We argue that PSR--BH binaries can serve as ready-made testing grounds for proposed resolutions to the black hole information paradox. We propose using timing signals from a pulsar beam passing through the region near a black hole event horizon as a probe of quantum gravitational effects. In particular, we demonstrate that fluctuations of the geometry outside a black hole lead to an increase in the measured root mean square deviation of the arrival times of pulsar pulses traveling near the horizon.  This allows for a clear observational test of the nonviolent nonlocality proposal for black hole information escape. For a series of pulses traversing the near-horizon region, this model predicts an rms in pulse arrival times of $\sim30\ \mu$s for a $3 M_\odot$ black hole, $\sim0.3\, $ms for a $30 M_\odot$ black hole, and $\sim40\, $s for Sgr A*. The current precision of pulse time-of-arrival measurements is sufficient to discern these rms fluctuations. This work is intended to motivate observational searches for PSR--BH systems as a means of testing models of quantum gravity.

\end{abstract}


\maketitle
\flushbottom

\section{Introduction}
Observable effects of quantum gravity are highly elusive.  One reason is that the Planck scale, where such effects are expected to be manifest, is well beyond the reach of current experiments.  In addition, regions of strong spacetime curvature, in which general relativity is expected to break down and be superseded by quantum gravity, tend to be hidden from view behind horizons. 

However, black holes may offer a unique low-energy window into quantum gravity.  The process of black hole formation and evaporation poses a fundamental challenge to conventional low-energy physics. If unitarity is not violated in black hole evolution, information stored inside the black hole must somehow emerge and the local quantum field theoretic description of a semi-classical event horizon must be significantly modified.  Quantum gravity will determine the way unitarity is ultimately preserved, and, if we are lucky, signals of this process may be detectable.

It is currently an open question how quantum gravity resolves the black hole information paradox.  Although most attempts to answer it have employed purely theoretical arguments, observational data may provide answers or at least constraints.  While some alternatives, such as the firewall scenario  \citep{Almheiri:2012rt}, predict phenomena that are extremely difficult to detect, other recent proposals feature large-scale nonlocal effects, e.g., \citet{Giddings:2004ud}, \citet{Papadodimas:2012aq}, \citet{Dvali:2012en}, \citet{Freidel:2015pka}, \citet{Hawking:2016msc}.  In particular, the nonviolent nonlocality proposal of \citet{Giddings:2011ks, Giddings:2012gc, Giddings:2014nla, Giddings:2014ova, Giddings:2016tla} has potentially observable consequences. Thus, the practical question is, where is the best place to hunt for these elusive signals of quantum gravity?

The Event Horizon Telescope (EHT) has generated substantial interest in the possibility of observing quantum gravitational effects in the near-horizon region of  Sagittarius A* (Sgr A*), the supermassive black hole at the center of the Milky Way galaxy. Direct observations of the accretion disk might exhibit modification of structures, such as the black hole shadow and the photon ring, predicted by general relativity \citep{Giddings:2014ova}.  

Another suggested method to investigate the near-horizon region, inspired by recent advanced Laser Interferometer Gravitational-wave Observatory (aLIGO) discoveries \citep{Abbott:2016blz, Abbott:2016nmj}, is to observe the gravitational waves emitted by the merger of two black holes.  One could hope to detect deviations in the signal from the general relativistic prediction \citep{Giddings:2016tla}.  The potential of gravitational wave astronomy has justifiably generated a great deal of excitement, but with so far only a handful of observed events so far, it is at best a promising but untested technique.  Furthermore, both the substantial observational noise and the theoretical uncertainties inherent in the difficult numerical modeling of the inspiral process limit the ability to discern the effects of quantum gravity.

Classically, the structure of spacetime is established by sending clock readings between observers on a spatial coordinate grid. To explore quantum mechanical effects on spacetime, we suggest using timing signals from the most precise natural clock---a pulsar---carried by light beams through the region near an event horizon.
Therefore, we propose that a pulsar-black hole (PSR--BH) binary is an ideal astrophysical system for observing quantum gravitational effects.  
The fact that such a system is in some sense ``clean," that is, not muddled by other astrophysical features, allowed for precision measurements to be taken of a novel gravitational phenomenon, gravitational waves \citep{1982ApJ...253..908T,2010ApJ...722.1030W, 2016arXiv160602744W}. It is this critical property of such binary systems that allows them to be used as effective probes of quantum gravity.

Pulsar--neutron star (PSR--NS) binaries are very clean systems whose orbital parameters can be measured with extreme accuracy, and for this reason, they have played a long and valuable role as precision astrophysical laboratories.  Most famously, observations of the so-called ``binary pulsar'' (PSR 1913+16) were the first to confirm the existence of gravitational waves \citep{1982ApJ...253..908T}. In addition, general relativistic effects, such as the Shapiro time delay of pulsar radiation passing through the gravitational potential well of the companion NS, have been measured, yielding precision tests of GR \citep{2006Sci...314...97K}. 

PSR--BH binaries have been called the ``holy grail of astrophysics" \citep{FaucherGiguere:2010bq} both because of their unique potential and also because no actual examples have yet been found.  However, the prospects for discovery are promising; large numbers of PSR--BH binaries are predicted to exist near the galactic center and should be readily detectable by both the Evolved Laser Interferometer Space Antenna (eLISA) \citep{AmaroSeoane:2012je} and the Square Kilometer Array (SKA) radio telescope \citep{FaucherGiguere:2010bq}. When they are eventually found, there are a variety of proposals to use PSR--BH binaries to investigate the gravitational properties of black holes \citep{2013ApJ...778..145N, Liu:2014uka, Rosa:2015hoa, 2016CQGra..33k3001J}, test extensions of general relativity \citep{Christian:2015smg, Yagi:2016jml}, and search for quantum gravitational effects associated with warped extra dimensions \citep{Simonetti:2010mk}.  Additionally, \citet{2014MNRAS.445.3370P} and \citet{Pen:2014jqa} have suggested that lensing of pulsar emission passing near the horizon could provide increased sensitivity, allowing for the observation of quantum gravitational effects.

For a pulsar orbiting a BH with an orbital plane that is seen edge-on, as the pulsar passes behind the BH, the radiation pulse travels through the near-horizon region.  Because the pulsar signal can be characterized with exceptional precision, there is the possibility to detect even subtle quantum gravity effects.  
In particular, fluctuations of the near-horizon geometry alter the null geodesics along which the photons of the pulse travel, modifying the Shapiro time delay for each pulse.  The effect can be observed as an increase in the root mean square (rms) variation in the arrival times of pulses at the telescope. By comparing pulse time-of-arrival (TOA) measurements when the pulsar is behind the BH to when it is in front, one can discern this increase in the rms due to these near-horizon quantum gravitational effects.

We find that horizon-scale fluctuations of the near-horizon geometry of the magnitude estimated by \citet{Giddings:2014ova} will be detectable. For a series of pulses passing near the horizon, this version of the nonviolent nonlocality scenario predicts an rms in pulse arrival times of $\sim30$~$\mu$s for $3 M_\odot$ BH, $\sim0.3\, $ms for a $30 M_\odot$ BH, and $\sim40\, $s for Sgr A*. Standard radio-pulse TOA measurements have the ability to detect these effects. Observations of the type suggested here represent a definitive test of the model of nonviolent nonlocality presented in \citet{Giddings:2014nla}.

This paper is organized as follows.  Sec.~\ref{sec:info_paradox} explains why quantum gravitational effects outside a BH horizon might be observable in the context of resolving the BH information paradox. Sec.~\ref{sec:BH-NS_Binaries} provides more background on binary pulsars, particularly PSR--BH binaries, and Sec.~\ref{sec:PSR-Sgr_A*} considers the observational possibilities of pulsars in orbit around Sgr A*.  Sec.~\ref{sec:time_delay} presents a theoretical calculation of the modification, due to metric fluctuations, of the TOA for pulses traveling near the BH, with a concrete example given in Sec.~\ref{sec:simpex}. Sec.~\ref{sec:Observables} explains how this effect can be observed.  Finally, Sec.~\ref{sec:Discussion} concludes with open questions and directions for future research.

\section{Background}
\label{sec:background}

\subsection{Quantum Gravitational Effects near the Event Horizon}
\label{sec:info_paradox}
According to our current understanding of general relativity and local quantum field theory, the process of black hole evaporation leads to a paradox:  since no causal signal can emerge from behind the horizon, information falling into a black hole seemingly cannot escape; however, this would imply that the information is irrevocably lost, even when the black hole has completely evaporated, suggesting pure states evolve into mixed states and a breakdown of unitary quantum mechanical evolution.

A variety of possible resolutions of this crisis have been proposed, and nearly all feature some sort of radical modification of our current semi-classical description of black holes. 
Most recent proposals preserve unitary quantum evolution and conjecture some novel mechanism to allow information to emerge from behind the horizon.  What, if any, are the observable consequences of these conjectured modifications?

The escape of information from behind a black hole horizon necessitates a significant nonlocal process.  Small violations of locality are not enough; subtle correlations between Hawking photons, for example, cannot account for the vast amount of information that needs to emerge from the black hole in order to preserve unitarity \citep{Page:1993wv, Almheiri:2012rt, Mathur:2005zp}. Instead, much more drastic modifications are needed, such as entirely replacing the geometry near the horizon with a firewall.  

One particularly interesting alternative scenario proposed by \citet{Giddings:2011ks, Giddings:2012gc} suggests that modifications of local quantum field theory appear as long-wavelength fluctuations set by the Schwarzschild scale rather than the Planck scale.  Note that for a macroscopic black hole of mass $M_{\rm BH}$, the Schwarzschild radius $R_{\rm s} = 2 G M_{\rm BH} / c^2$, which completely characterizes classical, non-rotating black holes, is many orders of magnitude larger than the Planck length $l_{\rm p} = \sqrt{\hbar G/c^3}$. These are strong but low-energy fluctuations, with a large amplitude and a mild impact; infalling observers, for example, still travel unharmed through the horizon region.  As a result, this proposal is termed ``nonviolent nonlocality." 

The preservation of unitarity puts a lower bound on the size of the predicted quantum fluctuations.  As the black hole evaporates, information must be emitted at roughly the same rate as energy, i.e., one qubit per Schwarzschild time $R_{\rm s}/c$ \citep{Giddings:2013kcj}.  In the nonviolent nonlocality scenario, the information transfer is parametrized as a nonlocal coupling between the interior black hole state and fields outside the horizon. Generically, such a coupling leads to enhanced, non-thermal Hawking radiation and a departure from the thermodynamic description of a black hole with an entropy given by the Bekenstein--Hawking formula \citep{Giddings:2011ks, Giddings:2012gc, Giddings:2013kcj}. 

However, we focus on a particular version of the proposal \citep{Giddings:2014nla} in which the standard black hole thermodynamics is preserved because the black hole interior couples universally via the exterior stress tensor, which in turn sources near-horizon metric perturbations, transferring information to the outgoing Hawking radiation. The necessary information emission rate requires this coupling be at least of the order of one, implying low-energy, order-one metric fluctuations \citep{Giddings:2013kcj, Giddings:2014ova}.

It was proposed in \citet{Giddings:2014ova, Giddings:2016tla} and \citet{Giddings:2016btb} that these fluctuations may lead to detectable effects in the accretion disk of Sgr A*, which might be observed by the EHT in the near future. Properly interpreting observations of this kind involves, in part, understanding the dynamics of the accretion disk as well as the structure of the interstellar medium at the galactic center. 
In searching for signals whose origins are of a quantum gravitational nature, it is imperative that other standard explanations be clearly eliminated.  
We argue for an additional astrophysical system, the PSR--BH binary, as a clear window through which to try to observe these novel effects.

\subsection{Binary pulsar systems}
\label{sec:BH-NS_Binaries}


One concern with proposing to observe these effects in a PSR--BH binary is the simple fact that such a system has yet to be discovered. However, there are reasons to be optimistic that PSR--BH binaries could be found in the near future.
It is likely that only a small fraction of the pulsars in our galaxy have so far been found. Recent estimates suggest that a small but significant number of these pulsars will belong to a PSR--BH system \citep{Lipunov:2005sv}. Many undiscovered pulsars likely lie near the galactic center, where conditions are conducive to the formation of PSR--BH binaries \citep{FaucherGiguere:2010bq}. The SKA is expected to be extremely effective at detecting a large number of the pulsars in our own galaxy and perhaps even in nearby galaxies \citep{Carilli:2004nx}. This will dramatically increase the likelihood of discovering a PSR--BH system.



PSR--BH binaries could also be discovered through the detection of their gravitational wave emission, with follow-up observations of their pulsar emission by radio telescopes. eLISA is expected to detect large numbers of binary systems within our galaxy. Such systems are expected to include PSR--BH binaries \citep{AmaroSeoane:2012je}. Moreover, eLISA is nominally sensitive to systems with orbital periods on the order of minutes, meaning the binary components are relatively close. As discussed in Sec.~\ref{sec:orientation}, the observation discussed here requires the pulsar beam to pass near the event horizon of the BH, which is more likely for a binary pair with a relatively small separation.

\subsection{Pulsars in Tight Binary Orbits around Sgr A*}
\label{sec:PSR-Sgr_A*}

A population of pulsars is predicted to exist at relatively small distances, i.e., $\lesssim4000$~AU, from the galactic center \citep{Pfahl:2003tf, Zhang:2014kva}. Moreover, it has been claimed that the $\gamma$-ray excess seen emanating from the galactic center is sourced by a population of millisecond pulsars \citep{PhysRevLett.116.051102}. This opens up the possibility that such pulsars could be used to probe near the horizon of Sgr A*. This scenario has recently been discussed in \citet{Iwata:2016ivt}. However, this approach would raise new challenges created by the long orbital period of such pulsars around Sgr A* and potential interference caused by the complex environment near Sgr A*. However, such an approach would have the virtue of a greatly enhanced rms of pulse arrival times, as discussed in Sec.~\ref{sec:TOA}. Moreover, observations of pulsars orbiting Sgr A* would complement and help confirm any results obtained through shorter timescale observations of stellar-mass PSR--BH binaries.

\section{Fluctuations in the TOA for Near-horizon Pulses}
\label{sec:time_delay}

In this section we explain how quantum gravitational fluctuations of the BH geometry modify the timing of pulses traveling through the near-horizon region.  
Although the rotation period of the pulsar is extremely stable, making pulsars superb natural clocks \citep{1984JApA....5..369B, 2011RvMP...83....1H}, the TOA of pulses is modulated by a variety of effects (e.g., orbital motion).\footnote{For a good reference on the subject of the pulsar timing, we refer the reader to \citet{LorimerKramerBook}.} 

In addition, because pulse arrival times can be measured with high precision and the gravitational field in a PSR--BH binary is strong, the TOA measurements are sensitive to a variety of general relativistic effects. One effect, which is dependent on fluctuations of the geometry near the BH and is therefore relevant here, is the Shapiro delay; this is the extra time the pulse takes to travel through the gravitational field between the pulsar and the Earth.

For a slowly varying gravitational potential $\phi$, the spacetime metric in isotropic coordinates can be written to leading order in $\phi$ as
\begin{equation}
ds^2 = - \left(1 + 2 \phi(\vec{x}) \right) c^2 dt^2 + \left(1 - 2 \phi(\vec{x}) \right) d\vec{x}^2 \ .
\end{equation}
The photons of the pulse follow a null geodesic.  Working to linear order in $\phi$, this null geodesic is defined by
\begin{equation}
c dt = \pm \left( 1 - 2 \phi(\vec{x}) \right) |d\vec{x}| \,.
\end{equation}
Integrating from the time of emission $t_{\text{em}}$ to the time of observation $t_{\text{obs}}$ along the trajectory $\rho(t)$ the photon travels, we obtain
\begin{equation}
c (t_{\text{obs}} - t_{\text{em}}) = \int^{\rho_{\text{obs}}}_{\rho_{\text{em}}}  d\rho \left(1 - 2 \phi(\rho)\right) \,.
\end{equation}
The second term in the integrand gives the Shapiro time delay $\Delta$, a modification compared with the flat-space result for the observed TOA:
\begin{equation}
\label{eq:Shapdel}
\Delta =  - \frac{2}{c} \int^{\rho_{\text{obs}}}_{\rho_{\text{em}}}  d\rho \, \phi(\rho) \,.
\end{equation}
%
This expression gives the time delay to leading order in $\phi$.  To this order, the integral is over the zeroth-order, straight-line, flat-space trajectory. Lensing due to $\phi$ changes the trajectory at first order, but affects the time delay only at second order \citep{Teyssandier:2008yx}.

The effects modifying the TOA of the pulses occur on a variety of timescales. Shifts that are constant as the pulsar orbits around its companion are not detectable; only variations in the pulse time of flight are measurable. Other corrections are modulated on the timescale of the pulsar's orbital period and provide detailed information allowing the entire orbit to be completely and precisely reconstructed.
 
The quantum gravitational fluctuations we are interested in observing occur on the timescale set by the size of the companion BH, the Schwarzschild time $R_{\rm s}/c$.  A pulse traveling through the near-horizon geometry encounters a modification with a characteristic length scale $R_{\rm s}$ of the Schwarzschild geometry.\footnote{Because the companion BH will likely be rotating, the background geometry is more accurately described by a Kerr geometry.  However, for the effect on the Shapiro time delay considered here, this distinction is irrelevant.}  Such a metric deformation causes an adjustment to the Shapiro time delay. Since $R_{\rm s}/c$ for a stellar-mass BH is several orders of magnitude smaller than the pulsar period, each pulse passes through a different, independent metric fluctuation. For the purposes of this section, it is sufficient to consider a pulse with an infinitesimal temporal width. Details of the TOA measurements, for actual pulses with finite temporal width, are discussed in Sec.~\ref{sec:TOA}.

We can parametrize the metric fluctuations in terms of a correction to the gravitational potential\footnote{Note that we do not consider the most general metric perturbation because we are focused only on the impact on the Shapiro time delay.} $\delta\phi$. The modified gravitational potential near a BH with Schwarzschild radius $R_{\rm s}$ is then
\begin{equation}
\phi(\vec{x}) = - \frac{R_{\rm s}}{2|\vec{x} - \vec{x}_{\text{BH}}|} + \delta\phi(\vec{x}) \,,
\end{equation}
As a result of these fluctuations, the Shapiro time delay obtains a correction
\begin{equation}
\label{eq:correctionShapdel}
\delta\Delta =  - \frac{2}{c} \int  d\rho \, \delta\phi(\rho) \,,
\end{equation}
where the integral is over the path of the pulse.  The fluctuation has support only in the near-horizon region, that is, only within about $R_{\rm s}$ of the horizon; farther away, $\delta\phi \approx 0$.  Rescaling the integration variable by $R_{\rm s}$, the correction (\ref{eq:correctionShapdel})  becomes
\begin{equation}
\label{eq:correctionShapdelrescaled}
\delta\Delta =  \frac{R_{\rm s}}{c} K \,.
\end{equation}
where $K =  -2\int  d(r/R_{\rm s}) \, \delta\phi(r)$ is the dimensionless strength of the fluctuation, integrated over an order-one range.  The correction $\delta\Delta$ scales as $R_{\rm s}/c$, as expected on dimensional grounds, since the Schwarzschild radius sets all the scales of the fluctuations.

The proportionality constant $K$ summarizes all the information about a given metric fluctuation, including the strength and the spatial profile.  If the fluctuation $\delta\phi$ were rapidly varying over the integration region, $K$ might be expected to approximately vanish.  However, $\delta\phi$ instead has a characteristic wavelength $R_{\rm s}$, so the integral is only over roughly a single oscillation.

For a series of pulses, each encountering an independent $\delta\Delta$, the rms fluctuation of the time delay will be
\begin{equation}
\label{eq:sigmadelta}
\sigma_\Delta \sim \kappa \frac{R_{\rm s}}{c}  \, ,
\end{equation}
where $\kappa = \sqrt{\langle K^2 \rangle}$  quantifies the strength of the metric fluctuations.

As discussed in Sec.~\ref{sec:info_paradox}, this version of the nonviolent nonlocality proposal requires the near-horizon metric fluctuations $\delta\phi$ to be order-one for there to be sufficient information transfer across the BH horizon.  Although the explicit form of $\delta\phi$ is unknown, this implies that $K$ and $\kappa$ are order-one numbers.  

The magnitude of $\delta\phi$ is not predicted to vary from one BH to another. In that case, $\kappa$ should be approximately the same for different PSR--BH binaries.  The resulting $\sigma_\Delta$ should then be linearly proportional to the mass of the BH. To provide an explicit estimate of the expected effect on TOA measurements, we take $\kappa \simeq 1$ for several BH masses, obtaining
\begin{eqnarray}
M_{\rm BH} &=& 3 M_{\odot}: \ \sigma_\Delta \simeq 3 \times 10^{-5} \, \text{s}\,, \cr
M_{\rm BH} &=& 30 M_{\odot}: \ \sigma_\Delta \simeq 3 \times 10^{-4} \, \text{s}\,, \cr
M_{\rm BH} &=& M_{\rm Sgr A*} \simeq 4\times10^6 M_{\odot}: \ \sigma_\Delta \simeq 40 \, \text{s} \,. 
\end{eqnarray}


We should note that the above calculation was performed in a linearized regime for both the BH background and the quantum fluctuations. This approximation allows us to obtain an analytic expression for the fluctuation of the time delay (\ref{eq:correctionShapdel}). However, the assumptions behind this approximation could break down in several ways.

If, for example, the pulsar is almost directly behind the BH, photons initially headed straight for Earth will be strongly lensed.  Null geodesics with a minimal isotropic radial coordinate less than $(1+\sqrt{3}/2) M_{\rm BH} = (2+\sqrt{3}) R_{\rm s} \equiv R_{\rm sh}$ are sufficiently curved that they fall into the horizon and thus are behind the BH shadow.  In order to reach Earth, photons must follow a curved trajectory just beyond the limits of the BH shadow.  The quantum fluctuations therefore must extend at least beyond the radius of the BH shadow $R_{\rm sh}$ to be detectable by pulsar pulses.  In that case, the expression (\ref{eq:correctionShapdel}) will be modified.  We still expect an rms fluctuation of the time delay qualitatively similar to (\ref{eq:sigmadelta}); however, a precise calculation involving numerical ray tracing is beyond the scope of this paper.

Alternatively, the linearized approximation could break down if the quantum gravitational fluctuations are very strong.  In that case, null geodesics connecting the pulsar with Earth would be highly lensed, and the flight time for photons traveling along such trajectories would be significantly modified. We expect that both of these effects lead to an enhanced rms fluctuation of pulse arrive times, which would be even more readily observable than the linearized rms fluctuation (\ref{eq:sigmadelta}).  We will address the effects of strong, nonlinear fluctuations in a future work.
With those limitations in mind, in Sec.~\ref{sec:TOA} we will discuss the observational consequences of these fluctuations in TOA measurements, and demonstrate that the $\sigma_\Delta$ resulting from $\kappa \simeq 1$ is expected to be detectable. However, in order to make the previous analysis more concrete, we first present an explicit calculation of $\sigma_\Delta$ due to a specific metric fluctuation ansatz.

\subsection{A Specific Example}
\label{sec:simpex}

The exact modification of the Shapiro time delay (\ref{eq:correctionShapdelrescaled}) depends on the details of the sources responsible for the transfer of information across the BH event horizon.  We illustrate the above discussion with an explicit but ad hoc example, and for simplicity we model the pulse as a single photon traversing near the near-horizon region.  

For the metric fluctuation, we consider the follow ansatz with a Gaussian radial profile, using spherical, isotropic coordinates centered around the black hole:
\begin{equation}
\label{eq:ansatz}
\delta\phi = \epsilon \, e^{-\left(r/N R_{\rm sh}\right)^2} \,,
\end{equation}
where $\epsilon$ is a dimensionless constant that represents the strength of the gravitational deformation due to nonlocal sources, following \citet{Giddings:2014ova}, $R_{\rm sh} = (1+\sqrt{3}/2)M_{\rm BH} = (2+\sqrt{3}) R_{\rm s}$ is the radius of the black hole shadow (in isotropic coordinates), and $N$ controls the distance from the BH over which the fluctuation extends.  As discussed in Sec.~\ref{sec:info_paradox}, it is expected that $\epsilon$ and $N$ are both order-one.  For this example calculation, we will assume $N > 1$. Otherwise, the fluctuations will not appreciably extend beyond the black hole shadow and will be undetectable by this method.

\begin{figure}[!ht]
    \centering
    \includegraphics[width=0.8 \textwidth]{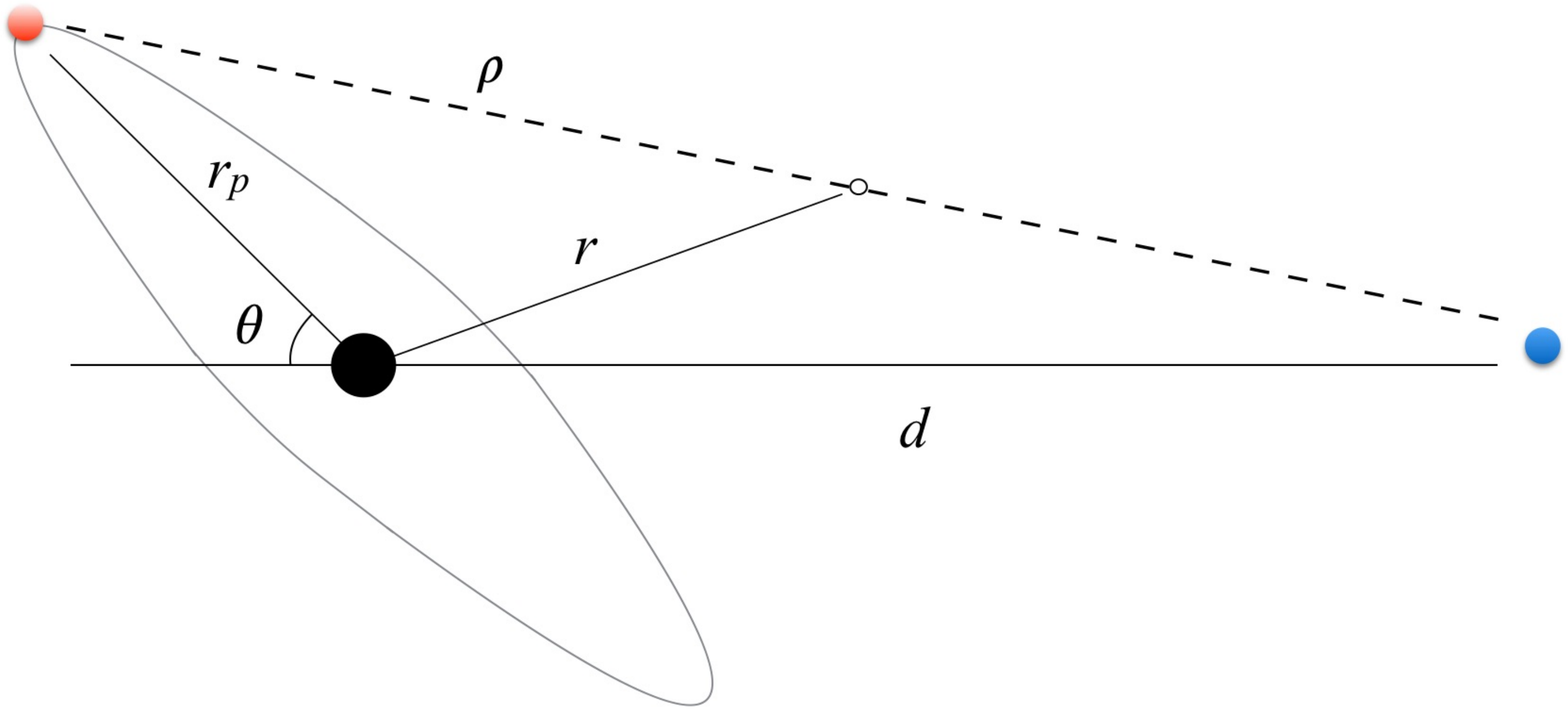}
    \caption{Depiction of a pulse traveling from a PSR--BH binary to Earth.  The blue dot is the Earth, the red dot is the pulsar, the black dot is the black hole, and the white dot is the photon.  The dashed line denotes the null geodesic followed by the pulse as it travels to the observer. Note that, to the order we are working, lensing of the pulse does not affect the time delay, and we may integrate over the straight, flat-space geodesic.} 
    \label{fig:setup}
\end{figure}

The geometry of the PSR--BH binary system as observed from Earth is illustrated in Fig.~\ref{fig:setup}.  We denote by $\theta$ the angle between the orbital plane of the binary and the line of sight.  A binary seen edge-on has $\theta = 0$, while $\theta = \pi/2$ denotes one seen face-on.\footnote{The usual inclination angle, the angle between the line of sight and the normal to the orbital plane, is $i = \pi/2 - \theta$.} The pulsar orbits at a distance $r_{\rm p}$ from the BH, and the binary is a distance $d$ from the Earth.  The binary is far from the Earth, so $r_{\rm p} \ll d$ and the lines of sight from the BH and from the pulsar are approximately parallel. The variable $\rho$ parametrizes the distance traveled by the photon away from the pulsar. A photon at a distance $\rho$ from the pulsar has a radial distance $r$ from the BH  given by $r^2 = \rho^2 + r_{\rm p}^2 - 2  r_{\rm p} \rho \cos\theta$.

The line integral (\ref{eq:correctionShapdel}) can be computed explicitly:
\begin{eqnarray}
\label{eq:deltaDeltaexponential}
\delta\Delta &=&  - \frac{2}{c} \int_{0}^{d} d\rho \ \delta\phi(\rho)  \nonumber\\
&=& - \frac{2}{c} \int_{0}^{d}d\rho \ \epsilon \exp\left[{-\frac{\rho^2 + r_{\rm p}^2 - 2 r_{\rm p} \rho \cos\theta}{(N R_{\rm sh})^2}}\right]  \nonumber\\
&=& - \epsilon \frac{\sqrt{\pi} N R_{\rm sh}}{c} e^{-\frac{r_{\rm p}^2 \sin^2\theta}{N^2 R_{\rm sh}^2}} \left\{\text{Erf}\left(\frac{r_{\rm p}}{N R_{\rm sh}} \cos\theta \right) + 1  \right\} \,.
\end{eqnarray}
The minimum distance between the photon and the BH is $r_{\rm p} \sin\theta$.  Due to the Gaussian factor in equation (\ref{eq:deltaDeltaexponential}), $\delta\Delta$ will only be significant when the photon comes within $NR_{\rm sh}$ of the BH.

The maximum effect occurs when $\theta = 0$, meaning the pulsar is directly behind the BH. Assuming the pulsar is outside the near-horizon region $r_{\rm p} \gg NR_{\rm s}$, we obtain
\begin{equation}
\label{eq:deltaDeltamax}
\delta\Delta_{\text{max}} = -2 \epsilon \sqrt{\pi} N R_{\rm sh}/c \,,
\end{equation}
and therefore, using $R_{\rm sh} = (2+\sqrt{3}) R_{\rm s}$, and comparing with (\ref{eq:correctionShapdelrescaled}), we find $K = -2 \epsilon \sqrt{\pi} (2+\sqrt{3}) N$.  

Using the metric fluctuation ansatz (\ref{eq:ansatz}), we computed the correction to the Shapiro time delay for a single pulse.  To obtain the rms fluctuation of the TOA measurements for a series of pulses, assume that each pulse passing near the BH encounters a $\delta\phi$ of the same magnitude and functional form but with a random sign; half of the pulses arrive early by an amount (\ref{eq:deltaDeltamax}), and the other half arrive late by the same amount. This gives $\kappa = 2 \epsilon \sqrt{\pi} (2+\sqrt{3}) N$.

This specific example illustrates the general argument presented above that the metric fluctuation amplitude $\epsilon$ and the resulting $\kappa$ have the same order of magnitude. If $\epsilon \simeq 1$ as predicted by the nonviolent nonlocality scenario, then we expect $\kappa \simeq 1$ as well.

\section{Observables}
\label{sec:Observables}

\subsection{TOA Measurements}
\label{sec:TOA}
As noted above in Sec.~\ref{sec:BH-NS_Binaries}, TOA measurements of pulses from pulsars have been used to probe general relativistic effects with high precision, as in the case of the binary pulsar \citep{1982ApJ...253..908T}. Such measurements could be used to detect the effects discussed in Sec.~\ref{sec:time_delay}. 

There is a well established methodology for making TOA measurements \citep{LorimerKramerBook}. The observations can account for every single rotation of the pulsar over very long periods of time. However, since the pulse shape changes from pulse to pulse, to make TOA measurements astronomers use the average pulse profile obtained for, say, five minutes of observations by ``folding'' (overlaying and averaging) the data on itself, modulo the pulse period. This average pulse profile is quite stable over very long periods of time (years to decades). The five-minute average pulse profile is temporally correlated with a model profile (e.g., what the average pulse profile looks like on long timescales), to compute the temporal offset of the five-minute average pulse with respect to the predicted arrival time. The specific pulse assigned the specific TOA is typically the pulse nearest the center of the observing window (e.g., the five-minute window).
Given the well-determined pulse period, one can predict the TOA for the next set of pulses. 

The precision of TOA measurements for any particular pulsar is given by
\begin{equation}
\sigma_{\rm TOA}\approx \frac{W}{S/N}\propto \frac{P\eta^{3/2}}{\sqrt{t_{\rm int}\Delta f}} \frac{S_{\rm sys}}{S_{\rm mean}}
\end{equation}
where $W$ is the pulse width, $S/N$ is the signal-to-noise of the average pulse profile, $P$ is the pulsar period, $\eta$ is the duty cycle ($\eta=W/P$), $t_{\rm int}$ is the integration time over which pulses will be averaged to create the average pulse profile, $\Delta f$ is the observing bandwidth, $S_{\rm sys}$ is the observing system's equivalent flux density, and $S_{\rm mean}$ is the flux density of the pulsar (average over $t_{\rm int}$). For $t_{\rm int}\sim5$~minutes, the rms precision of TOA measurements for a millisecond pulsar could be approximately $\sigma_{\rm TOA}\sim1\,\mu$s. High precision is best obtained in the case of millisecond pulsars, which also have the lowest ``timing noise" (i.e., a quasi-random walk in TOAs on timescales of months to years). A precision of 0.1\,$\mu$s was obtained by \citet{2001Natur.412..158V} for hour-long integrations using the Parkes 64m radio telescope on PSR J0437$-$4715, a bright millisecond pulsar in a WD-NS binary system. The rms precision scales as $1/\sqrt{t_{\rm int}}$, or $1/\sqrt{N_{\rm p}}$, where $N_{\rm p}$ is the number of pulses averaged. These fluctuations are the ``typical" rms fluctuations in TOA measurements---what would be observed if metric fluctuations were not present.

If the line of sight to a pulsar passes near the event horizon of the BH (i.e., the pulsar is ``behind'' the BH), fluctuations in the metric would increase the rms of the TOA measurements. A single pulse can be thought of as a series of photons, each delayed or advanced by the randomly fluctuating metric relative to the travel time in the absence of metric fluctuations. If the timescale for metric fluctuations is short compared to the pulse width $R_{\rm s}/c \ll W$, as in the case of a stellar-mass BH, then parts of each pulse of duration $R_{\rm s}/c$ are independently advanced or delayed by the metric fluctuations. The TOA measurement is usually dominated by the offset of a high signal-to-noise feature in the profile (e.g., a sharp leading edge), and the temporal advance or delay of that feature will be what is determined.\footnote{If, at the other extreme, $P\ll R_{\rm s}/c$, where $P$ is the pulse period (as is the case for a supermassive BH), then successive groups of pulses are independently advanced or delayed. We will discuss that case later in this section.} 

If $\sigma_\Delta$ is the rms delay or advance of a photon due to metric fluctuations and set by equation (\ref{eq:sigmadelta}), the resulting observed distribution of TOA measurements is the convolution of the distribution of TOA measurements in the absence of metric fluctuations (or for the pulsar ``in front" of the BH), which has a temporal width of $\sigma_{\rm TOA}$ with, we assume, an approximately Gaussian function of temporal width $\sigma_\Delta/\sqrt{N_{\rm p}}$. Thus the effect of metric fluctuations on the TOA measurements could be discerned if 
\begin{equation}
\frac{\sigma_\Delta}{\sqrt{N_{\rm p}}}\gtrsim\sigma_{\rm TOA} \ .
\end{equation} 
Note that increasing the duration of the observations, and thus increasing $N_{\rm p}$, will not help in measuring the effect of metric fluctuations, since both sides of the inequality scale as $1/\sqrt{N_{\rm p}}$. 

We assume for the purposes of discussion that $\sigma_{TOA}\sim$~1\,$\mu$s for $t_{\rm int}\sim5$~minutes. For a pulse period of $\sim 10\,$ms, we can detect $\sigma_\Delta\gtrsim0.1\,$ms, a value that is dependent only on the precision of the TOA pulse measurements and is independent of the BH mass. Since $\sigma_\Delta \sim \kappa R_{\rm s}/c$, this implies TOA observations can discern
\begin{equation}
\kappa\gtrsim 1 \left(\frac{M_{\rm BH}}{10\ M_\odot}\right)^{-1}
\end{equation}
where $M_{\rm BH}$ is the mass of the BH.  If we consider a $3 M_\odot$ BH, the most common mass for a stellar BH, we find the sensitivity to $\kappa\gtrsim 3$. For a $30 M_\odot$ BH, similar to the black holes in the binary system detected by aLIGO, we find the sensitivity to $\kappa \gtrsim 0.3$.

If we consider using pulsar emission to probe the event horizon of Sgr A*, then $R_{\rm s}/c\simeq80\,$s, which is longer than both the pulse width and pulsar period. In this situation, to calculate $N_{\rm p}$ we must divide the integration time by the Schwarzschild time instead of the pulsar period. Doing so and repeating the analysis outlined above implies that such a measurement could be used to probe a value of $\kappa\gtrsim 2\times10^{-8}$. Certain aspects of this approach are potentially problematic. The environment the pulse must travel through is more varied and complex than in the case of a binary system, and the orbital period of the pulsar will be much longer than in a standard PSR--BH binary. This may be compensated for by the dramatic increase in the rms in the pulse arrival times caused by the large size of the near-horizon region. In addition, \citet{2012ApJ...747....1L} suggest that high-frequency radio observations of pulsars in orbit around Sgr A* could alleviate some of these difficulties.

It is worth considering that, in principle, the precision of the TOA observations discussed above can be improved upon for extremely bright pulsars or more sensitive radio observatories. This could allow for a best-case sensitivity of perhaps an order of magnitude better than the nominal values quoted above. 

Finally, as noted in Sec.~\ref{sec:time_delay}, the observed rms TOA fluctuations caused by the metric fluctuations in equation (\ref{eq:sigmadelta}) scale linearly with the Schwarzschild radius and thus with the black hole mass.  Observations of multiple PSR--BH binary systems with different BH masses would therefore allow for this scaling to be observed and would serve as a means of distinguishing the effect predicted here from other phenomena.  

\subsection{PSR--BH Orientation}
\label{sec:orientation}
In order to have a chance to observe the effects of near-horizon metric perturbations, an appropriate PSR--BH system must first be discovered.  Such a binary must be observed sufficiently edge-on so that when the pulsar is behind the BH, the pulses pass close enough to the BH that they encounter the fluctuating geometry.  Using the notation of Sec.~\ref{sec:simpex}, for a PSR--BH system with an orbital plane inclined to the line of sight by an angle $\theta$ (where $\theta = 0$ corresponds to exactly on edge) and where the pulsar and BH are separated by a distance $r_{\rm p}$, the pulses pass within $r_{\rm p} \sin\theta$ of the BH; see Fig.~\ref{fig:setup}. Large-amplitude metric perturbations are expected to extend to an order-one number of Schwarzschild radii from the horizon.  Therefore, as we saw from equation (\ref{eq:deltaDeltaexponential}), for an appreciable modification of the Shapiro time delay, the PSR--BH binary must be oriented such that 
\begin{equation}
\label{eq:orientation}
\sin\theta \lesssim (2+\sqrt{3}) N R_{\rm s}/r_{\rm p} \ .
\end{equation}
The pulsar is expected to be far outside the near-horizon region, $r_{\rm p} \gg R_{\rm s}$, implying that $\theta \ll 1$.
Assuming that PSR--BH binaries are randomly oriented, the normal to the orbital plane is uniformly oriented in three-dimensional space. For $\theta\ll 1$, the solid angle for which equation (\ref{eq:orientation}) is satisfied is $2\pi (2+\sqrt{3}) N R_{\rm s}/r_{\rm p}$.  Thus, the probability that a given PSR--BH binary will be oriented sufficiently edge-on is $(1+\sqrt{3}/2) N R_{\rm s}/2r_{\rm p}$.

The orbital distance $r_{\rm p}$ is related via Kepler's third law to the orbital period of the system. To be a successful probe of the near-horizon region, the orbital period of a given PSR--BH system should be small enough to allow the pulsar beam to pass near the event horizon but not smaller than the rotational period of the pulsar, which would invalidate the observational approach outlined above. For very short orbital periods, the pulsar timing measurements would probably have to select data from successive orbital configurations (e.g., successive moments when the PSR is behind the BH) for folding.

To evaluate the likelihood of finding an appropriate PSR--BH binary, we consider a potential detection by eLISA. This gravitational wave detector will be most sensitive to systems with short orbital periods, on the order of minutes, but which are still large compared to a millisecond pulsar rotation rate.  Assuming follow-up radio observations could subsequently detect the pulsar emission, eLISA is well tuned to find the desired PSR--BH binaries.

For definiteness, we choose some nominal values in order to estimate the probability of detecting such a system, but it is important to note that there are great uncertainties in both the likely properties of such binaries and in their relative populations. For a PSR--BH binary with a $\sim 30M_\odot$ BH, having a Schwarzschild radius of $R_{\rm s}\simeq10^5\,$m, and an orbital period on the order of minutes, Kepler's third law yields an orbital radius $r_{\rm p}\simeq10^8\,$m. Although the value of $N$ is not well determined, an optimistic assumption of $N=10$ yields a probability of $\sim\frac{1}{30}$ that a given PSR--BH binary will be oriented sufficiently edge-on. If we consider the less optimistic value of $N=2$ we find a probability of $\sim\frac{1}{150}$. For a $\sim3M_\odot$ BH with the same set of parameters the probability is $\sim\frac{1}{300}$ for $N=10$ and $\sim\frac{1}{1500}$ for $N=2$. eLISA could detect tens of NS--BH binaries \citep{AmaroSeoane:2012je}. The pulsar emission from a significant number of these systems will be observable. These systems will necessarily each possess different orbital periods and black hole masses and thus the probability of performing the observation advocated here will be different for each. Nonetheless, the discovery of such systems represents a significant chance of locating one that could be used to search for  metric fluctuations near the event horizon. 

If we consider the potential for detection by SKA, the range of orbital periods for observed PSR--BH systems would be much larger, reducing the probability that, for a given binary, the pulsar beam would pass near the event horizon. However, this is compensated for in part by the large number of pulsars expected to be found. Using the results discussed in \citet{Lipunov:2005sv} and assuming $\sim10^5$ observable pulsars in our galaxy, SKA might detect $\sim100$ PSR--BH binaries. Depending upon the distribution of binary orbital radii, there could be a number of usable PSR--BH systems.

We also consider the likelihood of detecting a pulsar near Sgr A* oriented so that the pulsar beam passes near the event horizon. There are perhaps $\sim100$ pulsars surrounding Sgr A* with orbital periods $\lesssim10$ years \citep{Pfahl:2003tf}. Sgr A* has mass of $4\times10^6 M_{\odot}$ and  a Schwarzschild radius of $R_{\rm s}\simeq10^{10}\,$m.  A pulsar with an orbital period of a few years then has an orbital radius of $r_{\rm p}\simeq10^{13}\,$m. If we once again assume $N=10$, this yields a probability of $\sim\frac{1}{30}$, and for $N=2$ it is $\sim\frac{1}{150}$, that the orbit of a given pulsar will be oriented sufficiently edge-on. 

It is also worth noting that eLISA will be capable of detecting compact objects orbiting very near supermassive black holes \citep{AmaroSeoane:2012km}. If a pulsar is found by eLISA to be very close to Sgr A*, it would have a much greater likelihood of being oriented to send pulses through the near-horizon region.

Finally, with the advent of future instruments capable of more sensitive radio and gravitational wave observations, the probability of locating a pulsar capable of probing the event horizon of a BH will only increase.

\section{Discussion}
\label{sec:Discussion}

In this paper we have argued that PSR--BH binaries are ideally suited to probe metric fluctuations near the event horizons of black holes.  More specifically, we have demonstrated that a particular version of the nonviolent nonlocality proposal to resolve the BH information loss paradox predicts that these fluctuations are sufficiently large to generate an observable increase of the rms in pulsar TOA measurements, given by equation (\ref{eq:sigmadelta}). The required rate at which information must be released from the black hole implies  $\kappa$ is of the order of one. It is thus important to note that $\kappa$ is not, in the context of this proposal, a tunable parameter. Our results demonstrate that observations of a PSR--BH binary with the right properties can be used to probe values of $\kappa\simeq1$, thus making such observations a definitive test of this scenario. Moreover, this method can be used to test other models of quantum gravity that predict anomalous behavior of the metric near the horizon.

This current work could be extended in several ways. The relatively simple modification of the BH gravitational potential considered in Sec.~\ref{sec:simpex} could be made more detailed and sophisticated. In addition, numerical simulations could shed light on the precise effect such metric fluctuations would have on pulses traversing the near-horizon region.  For a typical PSR--BH binary in which the pulsar and BH are well separated, the modification of the Shapiro time delay is an observable consequence of the metric fluctuations predicted by the nonviolent nonlocality proposal.  However, if the pulsar itself travels through the near-horizon region, for example, in the final stages of a PSR--BH merger, other observable modifications to TOA measurements could arise. It is also worth considering ways in which PSR--BH binaries could be used to observe other quantum gravitational effects, such as enhanced Hawking flux or fluctuations in other metric components. 

Observational astronomy and theoretical work in quantum gravity have not traditionally had a great deal of overlap, except perhaps in a cosmological setting. However, Earth-based tests of quantum gravitational models are hard to pursue, while the universe provides ready-made astrophysical laboratories that can explore these extreme situations.  In addition to providing high-precision tests of classical general relativity and other relativistic gravity theories, pulsar--black hole binaries provide testing grounds for aspects of quantum gravity. We hope this paper will lead to further exploration of these possibilities and encourage the search for such laboratories in the sky.

\acknowledgments
We would like to thank Steve Giddings, Jonah Kanner, Jeffrey Kane, Steve Liebling, Andy O'Bannon, Peter Shawhan, Jamie Tsai, Joel Weisberg and I-Sheng Yang for their thoughtful comments and shared insights.

\bibliographystyle{aasjournal}

\end{document}